\documentclass[aps,pra,twocolumn,showpacs,preprintnumbers,superscriptaddress,floatfix,amsmath,amssymb,amsfonts]{revtex4-1}
\usepackage{graphicx}
\usepackage[bookmarks=true,colorlinks,linkcolor=blue,urlcolor=blue,citecolor=blue]{hyperref}
\usepackage{subfigure}
\usepackage{fourier}
\usepackage[T1]{fontenc}

\newcommand{\meV}[0]{\ensuremath{\mathrm{meV}}}
\newcommand{\eV}[0]{\ensuremath{\mathrm{eV}}}

\newcommand{\ket}[1]{\ensuremath\left\vert #1 \right\rangle}
\newcommand{\bracket}[1]{\ensuremath\left\langle #1 \right\rangle}

\newcommand{\mAA}[0]{\ensuremath{\mathrm{\AA}}}

\newcommand{\bise}{Bi${}_{2}$Se${}_{3}$}
\makeatletter
\@tfor\next:=abcdefghijklmnopqrstuvwxyzABCDEFGHIJKLMNOPQRSTUVWXYZ\do{%
  \def\command@factory#1{%
    \expandafter\def\csname bf#1\endcsname{\mathbf{#1}}
  }
 \expandafter\command@factory\next
}
\@tfor\next:=abcdefghijklmnopqrstuvwxyzABCDEFGHIJKLMNOPQRSTUVWXYZ\do{%
  \def\command@factory#1{%
    \expandafter\def\csname cal#1\endcsname{\mathcal{#1}}
  }
 \expandafter\command@factory\next
}
\makeatother
\renewcommand{\onlinecite}[1]{\cite{#1}}

\begin{document}

\title{Superconducting proximity effect in topological metals}
\author{Kyungmin Lee}
\author{Abolhassan Vaezi}
\affiliation{Department of Physics, Cornell University, Ithaca, New York 14853, USA}
\author{Mark H. Fischer}
\affiliation{Department of Condensed Matter Physics, Weizmann Institute of Science, Rehovot 76100, Israel}
\affiliation{Department of Physics, Cornell University, Ithaca, New York 14853, USA}
\author{Eun-Ah Kim}
\affiliation{Department of Physics, Cornell University, Ithaca, New York 14853, USA}
\pacs{71.10.Pm, 74.45.+c}

\begin{abstract}
Much interest in the superconducting proximity effect in three-dimensional (3D) topological insulators (TIs) has been driven by the potential to induce Majorana bound states at the interface.
Most candidate materials for 3D TI, however, are bulk metals, with bulk states at the Fermi level coexisting with well-defined surface states exhibiting spin-momentum locking.
In such topological metals, the proximity effect can differ qualitatively from that in TIs.
By studying a model topological metal-superconductor (TM-SC) heterostructure within the Bogoliubov-de Gennes formalism, we show that the pair amplitude reaches the naked surface, unlike in a topological insulator-superconductor (TI-SC) heterostructure where it is confined to the interface.
Furthermore, we predict vortex-bound-state spectra to contain a Majorana zero-mode localized at the naked surface, separated from the bulk vortex-bound-state spectra by a finite energy gap in such a TM-SC heterostructure.
These naked-surface-bound modes are amenable to experimental observation and manipulation, presenting advantages of TM-SC over TI-SC. 
\end{abstract}

\maketitle


\section{Introduction}


The potential realization of Majorana zero modes (MZMs) at the ends of a nanowire-superconductor hybrid system~\cite{MZF+12, DRM+12, RLF12, DYH+12, CMJ+13, FVHM+13} has attracted broad interest to different ways of stabilizing MZMs.
While there are proposals to exploit exotic statistics of MZMs within quasi-one-dimensional networks~\cite{STL+10, LSDS10, ORvO10, QHZ10}, a two dimensional setting would be desirable for observing statistical properties of MZMs.
A MZM can appear as a vortex bound state of triplet superfluids~\cite{KS91} or superconductors~\cite{RS95}.
Unfortunately, naturally occurring triplet superconductors are rare, and hence the proposal  by Fu and Kane~\cite{FK08} to use the superconducting proximity effect on the topological insulator (TI) surface states raised enthusiasm as an alternative route to realizing MZMs hosted in a two dimensional space.
However, most known three-dimensional (3D) TI candidate materials, such as \bise{} and Bi${}_{2}$Te${}_{3}$, have both the surface states and the bulk states at the Fermi energy~\cite{WXX+10}.
Recent experimental successes in inducing superconductivity in \bise{} thin films through proximity effect~\cite{WLX+12,WDF+13} makes it all the more urgent to address the superconducting proximity effect in such topological {\it metals}, where surface states and bulk states coexist.


In the proposal by Fu and Kane~\cite{FK08} for realizing MZMs, superconductivity is induced to the surface states of a 3D TI by proximity to a trivial $s$-wave superconductor (SC).
The argument for the existence of a MZM as a vortex bound state is based on the formal equivalence between a $p+ip$ superconducting gap of a spinless fermion and a trivial $s$-wave gap after projection to the space of surface states.
However, with only the surface states available at the Fermi energy, the superconducting proximity effect is limited to the interface between the TI and the adjacent superconductor.
On the other hand when the bulk band crosses the Fermi energy, as they do in many 3D TI materials, there is a chance that the proximity effect can reach the naked surface.
The key questions then would be (1) when can proximity effect reach the naked surface and (2) whether the naked surface can host MZMs.
These questions are the focus of this paper. 

\section{Model Hamiltonian for Heterostructure}

\begin{figure}
\centering
\subfigure[]{\label{fig:heterostructure}\includegraphics[width=0.35\columnwidth]{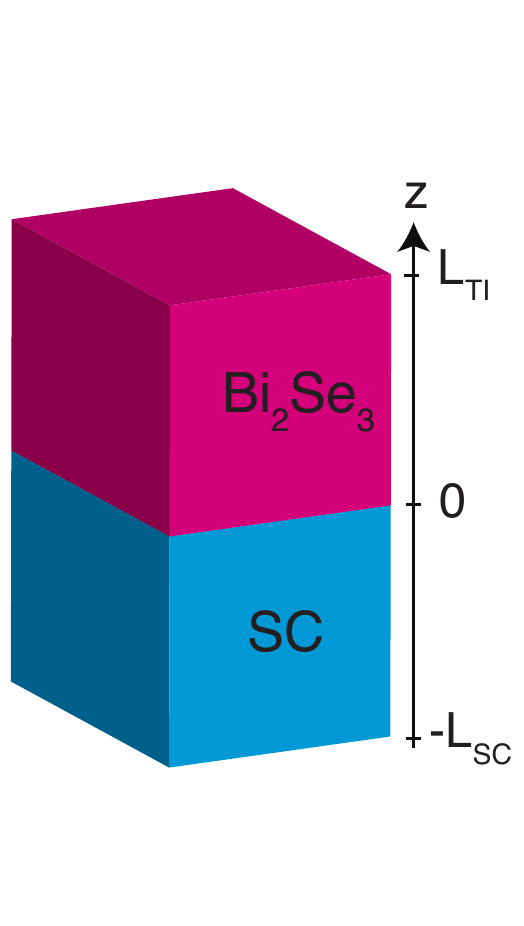}}
\subfigure[]{\label{fig:bandstructure}\includegraphics[width=0.55\columnwidth]{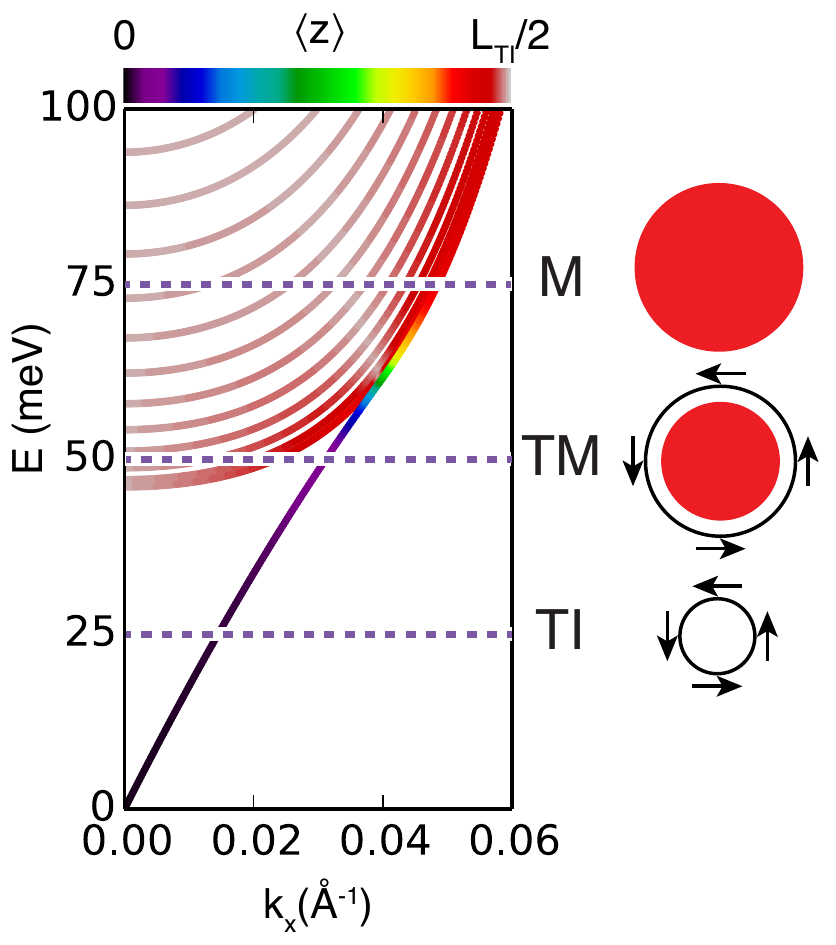}}
\caption{\label{fig:system}
(a) \bise{}-SC heterostructure considered in this paper.
(b) Dispersion of \bise{} on a slab of finite thickness $L_{\mathrm{TI}}$.
Each point is doubly degenerate, and the color scale indicates the minimum $z_{\mathrm{min}}=\min_{\Psi}\bracket{z}_{\Psi}$ that can be obtained within the degenerate space $\Psi \in \mathrm{span} \{\Psi_1,\Psi_2\}$.
The dotted horizontal lines indicate representative chemical potentials associated with TI, TM, and M regimes as defined in the text.
We present schematics of corresponding Fermi surfaces next to each dotted line,
where red filled circles represent the bulk states and the black circles the surface states.
Each arrow points along the direction of the spin of the surface state on one of the surfaces, which is locked to the momentum.
}
\end{figure}

To be concrete, we consider a \bise{}-SC heterostructure, where the \bise{}  takes the form of a finite thickness slab, so that we can study its naked surface [Fig.~\ref{fig:heterostructure}].
We first study how the proximity effect propagates differently depending on  the location of the chemical potential, by solving the Bogoliubov-de Gennes (BdG) equation in the heterostructure.
We then study the vortex bound state spectra with the gap structure inferred from the solution and investigate the stability of a MZM on the naked surface depending on chemical potential.

\begin{figure}
\centering
\subfigure[\label{fig:pairamplitude-a}]{\includegraphics[height=0.36\columnwidth]{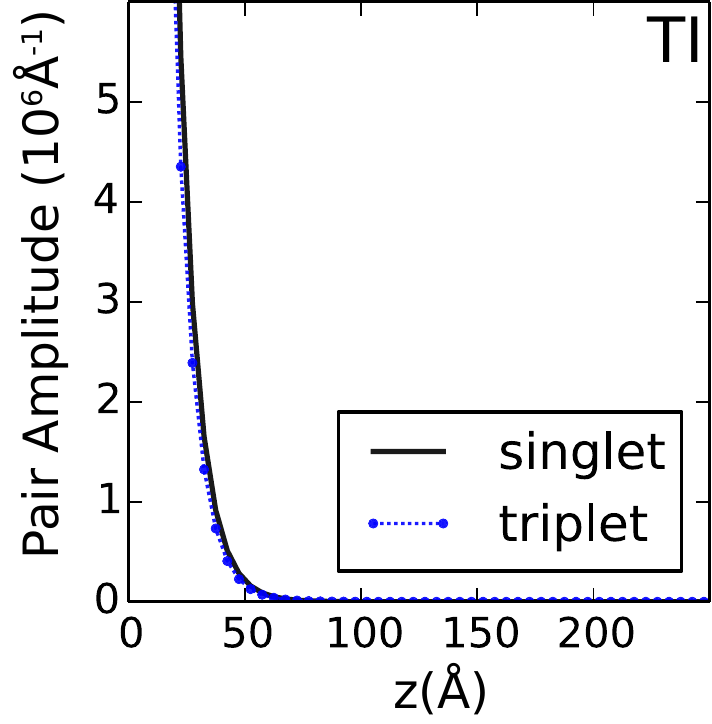}}
\subfigure[\label{fig:pairamplitude-b}]{\includegraphics[height=0.36\columnwidth]{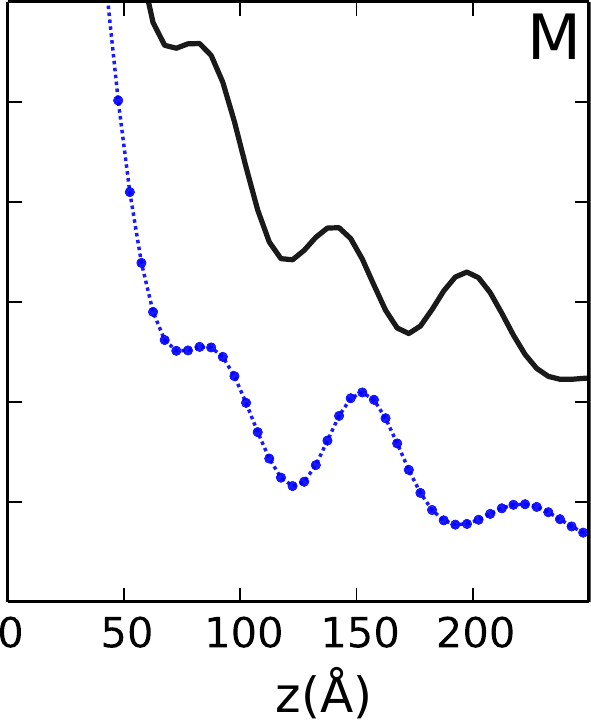}}
\subfigure[\label{fig:pairamplitude-c}]{\includegraphics[height=0.36\columnwidth]{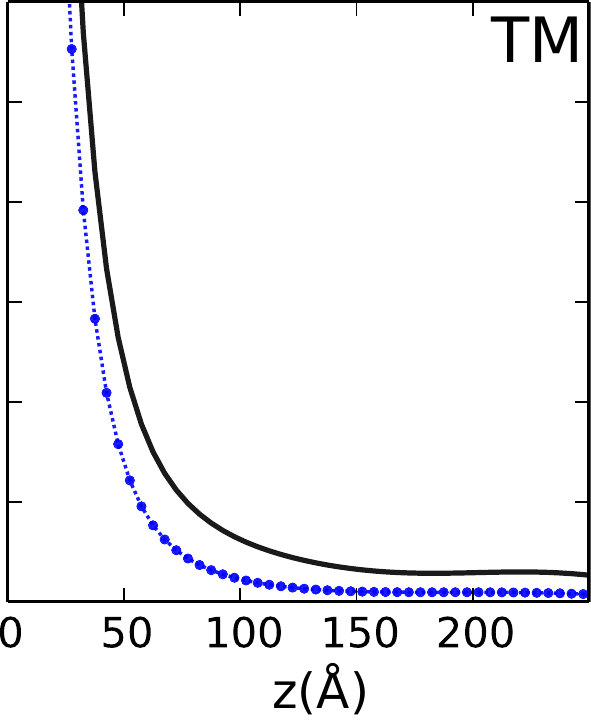}}
\caption{\label{fig:pairamplitude}
The pair amplitudes in singlet and triplet channels as a function of the distance from the interface boundary ($z$) in three regimes: 
(a) TI,
(b) M, and 
(c) TM,
with chemical potentials $\mu_{\mathrm{TI}} = 25\;\meV$, $\mu_{\rm M}=75\;\meV$, and $\mu_{\rm TM} = 50\;\meV$, respectively.
The parameters used in the calculation are
$L_{\mathrm{TI}} = 500\;\mAA$,
$L_{\mathrm{SC}} = 250\;\mAA$,
$a=5 \;\mAA$,
$\Delta_{0} = 5\;\meV$, 
$\mu_{\mathrm{SC}} = 300\;\meV$,
and with $k$ points on a $100\times100$ grid.
(One quintuple layer is roughly 10\AA.)
}
\end{figure}


The heterostructure of interest consists of a slab of \bise{} for $0<z<L_{\mathrm{TI}}$ and superconductor for $-L_{\mathrm{SC}} < z < 0$.
The electronic structure of \bise{} is described by an effective two-orbital Hamiltonian on a simple cubic lattice with lattice constant $a$.
Given the slab geometry with periodic boundary conditions in the $x$ and $y$ directions, we choose as basis $\ket{\bfk, z, \alpha, s}$, a state with momentum $\bfk=(k_x, k_y)$ within an $xy$ plane at $z=(n_z+1/2) a$ for $n_z=0 \ldots N_{\mathrm{TI}}-1$, with orbital $\alpha$ and spin $s$.
As the normal-state Hamiltonian of the model we take a lattice version of the four-band continuum model for 3D TI as given in Ref.~\onlinecite{LQZ+10} consisting of two parts: intra-layer terms $\hat{H}_{\bfk}^{0}$ and the inter-layer hopping (from $n_z$ to $n_z+1$) terms $\hat{H}_{\bfk}^{(1)}$ written as
\begin{align}
  \hat{H}_\bfk^{(0)}
    =& 
        t_0 - \mu - 2 t_1 \cos (k_x a) - 2 t_1 \cos (k_y a)
       \nonumber\\
     &
      + 
      \left[
        m_0 - 2 m_1 \cos (k_x a) - 2 m_1 \cos (k_y a)
      \right]
      \hat{\tau}_z \nonumber\\
     &
      +
      \lambda \sin (k_y a) \hat{\tau}_x \hat{\sigma}_x -
      \lambda \sin (k_x a) \hat{\tau}_x \hat{\sigma}_y 
     \nonumber
        \\
  \hat{H}_\bfk^{(1)}
    =&
     - t_2
     - m_2 \hat{\tau}_z
     - i \frac{\lambda'}{2} \hat{\tau}_y 
     \label{eq:hamiltonian}
\end{align}%
where $\hat{\tau}_i$($\hat{\sigma}_i$) for $i=x,y,z$ are Pauli matrices in the orbital (spin) space.
The parameters of the Hamiltonian in Eq.~\eqref{eq:hamiltonian} are chosen such that the model matches the continuum model for \bise{} from Ref.~\onlinecite{LQZ+10} up to $O(k^2)$ for $a=5\;\mAA$:
$t_1=1.216$ eV,
$t_2=0.230$ eV,
$m_0=7.389$ eV,
$m_1=1.780$ eV,
$m_2=0.274$ eV,
$\lambda = 0.666$ eV, and
$\lambda' = 0.452$ eV.
The reference chemical potential $t_0=5.089$eV has been chosen such that the degeneracy point of the surface state branch lies at $E=0$ when $\mu=0$.


To explicitly define what we mean by a topological metal (TM) it is important to recall the well-known band structure of the above model.
As shown in Fig.~\ref{fig:bandstructure}, the spectrum of the Hamiltonian contains a (degenerate) gapless branch in addition to the bulk states separated by a finite gap.
Depending on the chemical potential, we now define three regimes: topological insulator (TI), TM, and metal (M).
The TI is a bulk insulating state with the chemical potential within the bulk band gap [Fig.~\ref{fig:bandstructure}, $\mu =25$meV].
In the TI regime, gapless states at the Fermi level are highly localized at the two surfaces of the slab.
On the other hand, when the chemical potential is well within the bulk conduction band, all the states at the Fermi level, including the ones from the branch that contains surface states in the TI regime, are extended over the entire slab [Fig.~\ref{fig:bandstructure}, $\mu=75$meV].
Here, we refer to this regime as metal (M).
In between these two regimes, there is a range of chemical potential where 
the branch that is an extension of the Dirac cone coexists with the bulk states at the Fermi level, but nevertheless it remains surface-localized and spin-momentum locked [Fig.~\ref{fig:bandstructure}, $\mu=50$meV].
Experimentally, this regime can be identified through the spin-momentum locking of Dirac-cone states outside the bulk band-gap, which has been observed in \bise{} by spin-angle-resolved photoemission spectroscopy (ARPES)~\cite{HXQ+09}.
We refer to this regime as topological metal~\cite{Kar11,BBK+12,MR13,JKB+13,HFH+14}.
Note that while the existence of the in-gap surface states is protected by topology, its dispersion depends on material specific details.
Therefore, the exact ranges of chemical potential of the three regimes will also be material dependent. Nevertheless, the surface states and the bulk states have qualitatively different contributions to the proximity effect as we will see below,
and therefore we expect the three regimes in a real material to show qualitatively the same features as the corresponding regimes in our calculation.


For the superconductor part ($z<0$) we again use a two-orbital model of the same form as Eq.~\eqref{eq:hamiltonian} to describe its normal state, with $z=(n_z+1/2)a$ for $n_z = -N_{\mathrm{SC}}, \ldots, -1$.
The same parameters as \bise{} are used, except that we flip the sign of the ``mass term'' ($m_0 - 4 m_1 - 2 m_2$) and make the resulting band structure trivial, by choosing $m_0=7.949\;\eV$.
Also, since the inter-layer hopping in both parts of the heterostructure is described by the same term $\hat{H}_{\bfk}^{(1)}$, we use it to describe the tunneling between the two parts.

\section{Distance Dependence of Pair Amplitudes}


In order to compare the proximity effect in the three regimes, 
we impose an orbital-independent $s$-wave superconducting gap of strength $\Delta_0$ on the superconductor ($z<0$) and diagonalize the BdG Hamiltonian.
We then study how the resulting pair amplitude depends on the distance from the interface.
Because the pair amplitude is a matrix in both the spin and the orbital basis, it is convenient to look at its projection onto different spin channels.
As pointed out in Ref.~\onlinecite{BB13}, spin-singlet $A_{1g}$ pairing term induces spin-singlet $A_{1g}$ and spin-triplet $A_{2u}$ components of the pair amplitude matrix in the presence of spin-orbit coupling of the form Eq.~\eqref{eq:hamiltonian}.
The spin singlet and triplet components $\hat{F}^s(z)$ and $\hat{F}^t(z)$ are themselves $2\times2$ matrices in the orbital space, given by
\begin{align}
  \hat{F}_{\alpha \beta}^{\mathrm{s/t}}(z)
  &=
  \frac{1}{N}
    {\sum_{\bfk s_1 s_2}}
    \left[ \hat{S}^{\mathrm{s/t}}_\bfk \cdot i\hat{\sigma}_y \right]_{s_1 s_2}
      u_{\bfk z \alpha s_1}^{\phantom{*}}
      v_{\bfk z \beta s_2}^*,
\label{eq:F}
\end{align}
where $N$ is the number of $k$ points in the $xy$-plane and the sum is over every positive-energy BdG eigenstate $(u_{\bfk z \alpha s}, v_{\bfk z \alpha s})$.
In Eq.~\eqref{eq:F} $\hat{S}^{\mathrm{s}}_{\bfk}$ and $\hat{S}^{\mathrm{t}}_{\bfk}$ are the respective form factors for spin-singlet and triplet defined by
\begin{align}
\hat{S}^{\mathrm{s}}_{\bfk}
&= \hat{\sigma}_0
  ,\\
\hat{S}^{\mathrm{t}}_{\bfk}
  &=
  \frac{\sin (k_y a) \hat{\sigma}_x - \sin (k_x a) \hat{\sigma}_y}
         {\sqrt{\sin^2 (k_x a) + \sin^2 (k_y a)} },
\end{align}
with $\hat{\sigma_0}$ the ($2\times2$) identity matrix.
In the self-consistent approach with attractive interaction $U$ in the BCS channel, the superconducting gap $\Delta$ is proportional to the pair amplitude ($\Delta \sim U F$).
Here, however, no such self-consistency is imposed, and the pair amplitude inside the \bise{} is completely due to the Andreev reflection from the interface~\cite{And64,BTK82}.


We study the $z$-dependence of the pair amplitudes in \bise{} side ($z>0$) in the three regimes: TI, M, and TM.
For this purpose, we pick for each $z$ in each spin channel the largest eigenvalue $F^{\mathrm{s/t}}_{+}(z)$ of the $2 \times 2$ matrix $\hat{F}^{\mathrm{s/t}}(z)$, which indicates the leading instability in the given spin channel.
In all three regimes, both spin-singlet and spin-triplet pair amplitudes are expected to be non-zero because of the spin-orbit coupling term in the Hamiltonian~\eqref{eq:hamiltonian}.


In Fig.~\ref{fig:pairamplitude}, we plot $F^{\mathrm{s/t}}_{+}(z)$ as a function of $z$.
In the TI regime [Fig.~\ref{fig:pairamplitude-a}], we find that the pair amplitude is confined to the buried interface with exponential decay, since it is carried entirely by the surface states with such spatial profile.
In addition, singlet and triplet components of the pair amplitude have the same magnitude as a result of spin-momentum locking of the surface states.
In the M regime [Fig.~\ref{fig:pairamplitude-b}], on the other hand, the pair amplitudes show Friedel oscillations with an envelop that decays algebraically as a function of $z$.
(See the Supplemental Material for an analytic understanding of the $z$ dependence of the pair amplitudes in the M regime. \footnote{See Supplemental Material for an analytic derivation of the pair amplitude in a one-dimensional example model.})
In addition, the singlet channel dominates over the triplet channel in the M regime.


The results in the TM regime [Fig.~\ref{fig:pairamplitude-b}] can be understood by combining the two pictures of the TI and the M regimes.
In the TM regime, the pair amplitude consists of two components: the surface-states contribution and the bulk-states contribution, each of which should be qualitatively the same as the pair amplitude in the TI and the M regimes, respectively.
At large distances where the bulk-states contribution is dominant, the pair amplitude should show a power-law-like decay.
Friedel oscillation should also be present in principle, but in Fig.~\ref{fig:pairamplitude-c}, the large wavelength of the oscillation makes it difficult to identify the oscillation.
With the power-law decay of the pair amplitude at large distances, superconductivity can be induced on the naked surface by proximity effect in the TM.
This induced pairing on the naked surface is a mixture of singlet and triplet components.
The two components, however, lead to the identical effective BdG Hamiltonian for the surface states, as the surface states are fully spin-momentum locked.


\section{Majorana Vortex Bound State on the Naked Surface}

\begin{figure}
\begin{center}
\subfigure[]{\label{fig:vbs_ti_gap}\includegraphics[height=0.25\columnwidth]{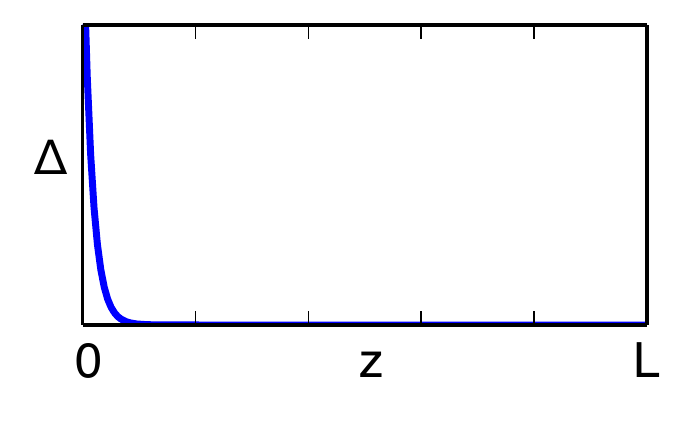}} %
\subfigure[]{\label{fig:vbs_tm_gap}\includegraphics[height=0.25\columnwidth]{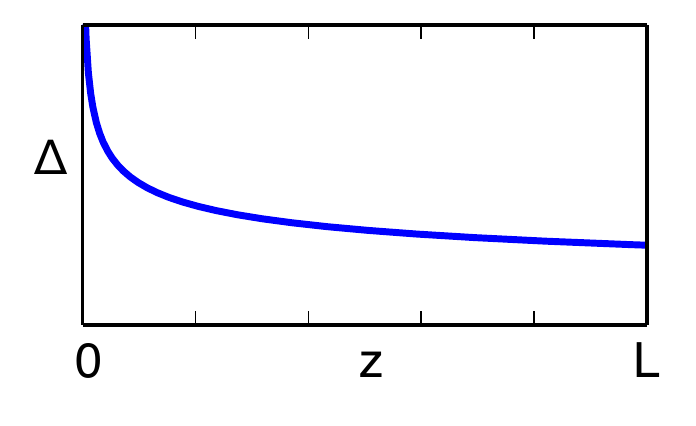}}
\phantom{XX}\\
\subfigure[]{\label{fig:vbs_ti_lowest}\includegraphics[height=0.4\columnwidth]{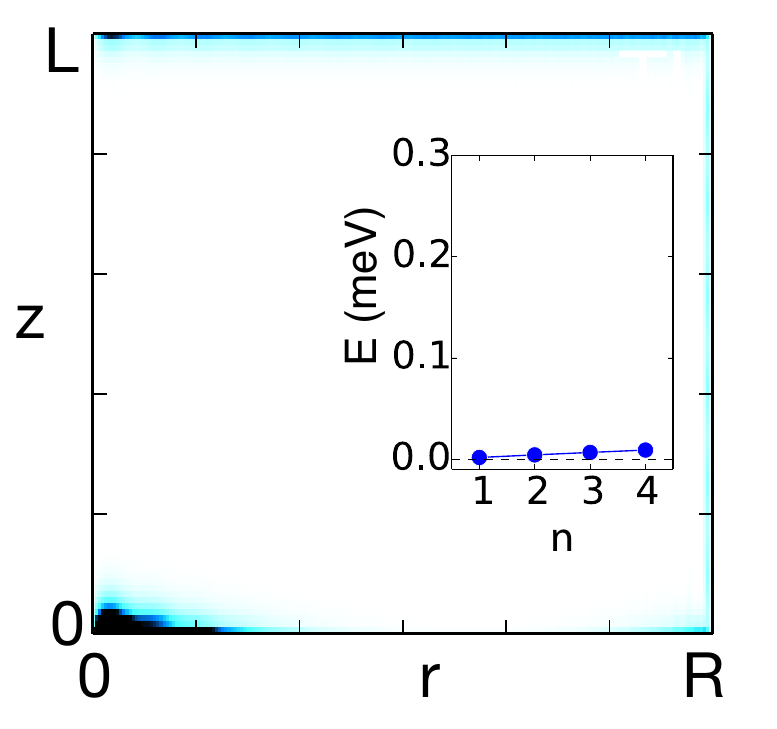}} %
\phantom{X}
\subfigure[]{\label{fig:vbs_tm_lowest}\includegraphics[height=0.4\columnwidth]{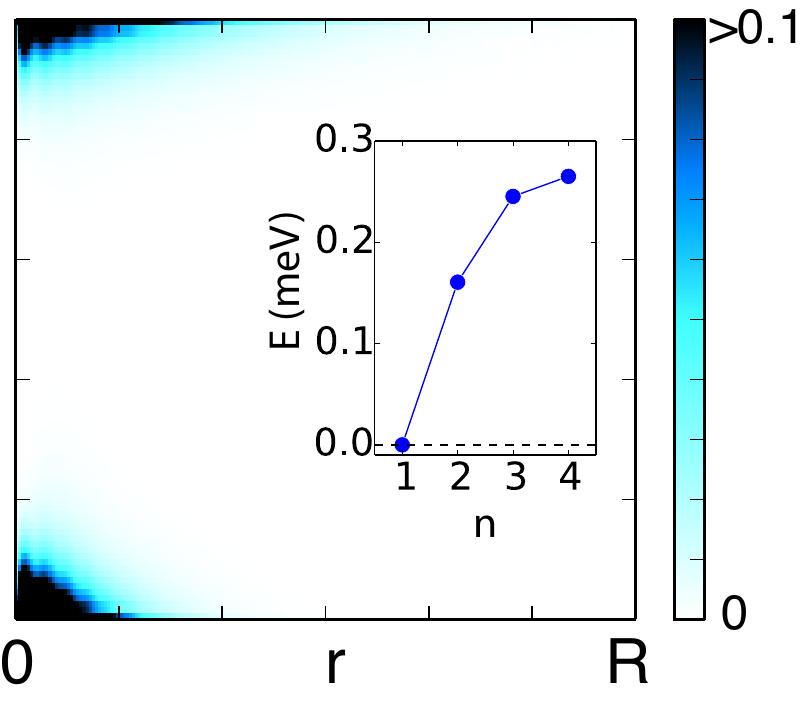}}
\caption{\label{fig:vortex}
Panels (a) and (b) show the $z$ dependence of the gap profile used to compute vortex-bound-state spectra for TI ($\mu=25\;\mathrm{meV}$) and TM ($\mu=50\;\mathrm{meV}$) regimes, respectively.
Panels (c) and (d) show the spatial probability density profile $\rho_n(r,z)$, as defined in Eq.~\eqref{eq:rho}, of the lowest lying vortex bound state in two regimes.
$\rho_n(r,z)$ has been normalized such that the maximum value is unity.
The parameters used in the calculation are
$a=5\;\mathrm{\AA}$,
$R=3000\;\mathrm{\AA}$,
$L=500\;\mathrm{\AA}$,
$\Delta_0 = 5\;\meV$,
$z_0 = a/2$,
$\xi_R = 100\;\mathrm{\AA}$,
and $\xi_L = 8\;\mathrm{\AA}$ for TI and $\gamma = 1/4$ for TM.
The inset in each case shows the vortex bound state spectrum, i.e. the energy $E_n$ of the $n$th excitation.
}
\end{center}
\end{figure}


Next, we ask whether the naked surface of a TM with proximity-induced superconductivity can host MZMs.
Formally related to the system of our interest is the 3D bulk superconducting Cu-doped \bise{}.
For this system Hosur {\it et al.}~\cite{HGMV11} predicted a vortex parallel to the $c$-axis to host a surface MZM even when the chemical potential is within the bulk conduction band, as long as it is below a critical value of $\sim 0.24\;\mathrm{eV}$ from the bottom of the band.
The chemical potential of an undoped \bise{} falls within this range~\cite{XQH+09}, and so does our definition of TM in our model.
Hence a vortex in a TM proximity-coupled to a superconductor is likely to host a protected MZM at the naked surface. However, the effect of $z$-axis-dependent proximity-induced pairing strength on the naked surface and energetic stability of the MZM are not known \textit{a priori}.


For concreteness, we solve the BdG equation on a cylindrical slab of \bise{} with thickness $L$ and radius $R$, with chemical potential in the TI and TM regimes.
With the axis of the cylinder aligned along the $z$ axis, we take the $xy$-coordinates to be continuous, while keeping the $z$ coordinate discrete.
The normal state Hamiltonian is then described by Eq.~\eqref{eq:hamiltonian}, with $\sin (k_i a) \rightarrow -i a \partial_i$ and $\cos (k_i a ) \rightarrow 1 + \frac{1}{2} a^2 \partial_i^2$ for $i=x,y$.
Informed by our proximity effect calculation above, we impose an $s$-wave superconducting gap of the following respective profiles for TI and TM:
\begin{align}
\Delta_{\mathrm{TI}}(r, \theta, z)
  &=
    \Delta_0 \tanh ( r/\xi_R ) e^{i \theta}
             e^{-(z-z_0)/\xi_z} \label{eq:deltati},\\
\Delta_{\mathrm{TM}}(r, \theta, z)
  &=
    \Delta_0 \tanh ( r/\xi_R ) e^{i \theta}
             \left(z/z_0\right)^{-\gamma},\label{eq:deltatm}
\end{align}
where $(r,\theta,z)$ is the cylindrical coordinate of the system.
$\xi_R$ and $\xi_z$ are superconducting correlation lengths in the radial and the axial directions, respectively.
We chose $z_0$ such that the bottom-most layer ($z=z_0$) of the TI/TM has a gap of magnitude $\Delta_0$, and a positive exponent $\gamma$ is used for the gap profile to decay as $z$ increases.


Because of the rotation symmetry of the system, it is convenient to use as basis the circular harmonics
\begin{align}
  \varphi_{\nu m} (r, \theta)
    &= 
      \frac{1}{\sqrt{\pi} R}
      \frac{ J_\nu ( \alpha_{\nu m}\,r/R ) }{J_{\nu+1} (\alpha_{\nu m})}
      e^{i \nu \theta},
\end{align}
where $J_{\nu}$ is the Bessel function of the first kind of order $\nu$ and $\alpha_{\nu m}$ is its $m$th zero.
Expressed in terms of
$\{ \varphi_{\nu m}\}$, the Hamiltonian can be block diagonalized into different sectors of $L_z + S_z + Q/2$, where $L_z$ and $S_z$ are orbital angular momentum and spin of a quasiparticle in the $z$ direction, and $Q$ is its charge in units of $|e|$ ($-1$ for electron).


One can then diagonalize each block of the Hamiltonian, and find the low energy eigenstates.
Each eigenstate $(u_{\alpha\sigma}^{n}(r,\theta, z), v_{\alpha\sigma}^{n}(r,\theta, z))$ can be identified using its spatial probability density defined as
\begin{align}
\label{eq:rho}
\rho_{n}(r,z)
  &\equiv  r
    \sum_{\alpha, \sigma}
    \int \frac{d \theta}{2\pi}
      |u_{\alpha\sigma}^{n}(r,\theta, z)|^2 + |v_{\alpha\sigma}^{n}(r,\theta, z)|^2.
\end{align}
Figures~\ref{fig:vbs_ti_lowest} and \ref{fig:vbs_tm_lowest} show $\rho_n(r,z)$ of the lowest excitation in the TI and TM regimes.
In the TI regime, the superconducting gap decays exponentially away from the bottom surface, becoming negligible on the top surface.
As a result a zero-energy vortex bound state appears only on the bottom surface, and the top surface remains metallic~[Fig.~\ref{fig:vbs_ti_lowest}].
The resulting spectrum is shown in the inset of Fig.~\ref{fig:vbs_ti_lowest}.
In the TM regime, on the other hand, the superconducting gap at the top surface is sizable, and a well-defined Majorana vortex bound state exists on both the top and the bottom surfaces.
Hence the TM regime brings the best of both worlds: a stable zero mode on the experimentally accessible top surface.
\footnote{The same calculation in the M regime with gap function given by Eq.~\ref{eq:deltatm} trivially yields no zero mode, as expected.}


\section{Conclusions}


In summary, we studied the proximity effect in \emph{topological metals}, i.e.,  topological insulators with bulk states at the Fermi level coexisting with well-defined surface states exhibiting spin-momentum locking.
Against the common belief that ideal topological insulators should be bulk insulating, we showed that the existence of bulk carriers can be a feature for the proximity effect as the induced gap will be observable at the naked surface.
Most importantly, we showed that a vortex line in a TM-SC structure will host an energetically stable Majorana bound state at the naked surface.


Although we focused on the proximity effect due to an $s$-wave superconductor for concreteness, our results are applicable to the proximity effect due to a $d$-wave superconductor such as the high-$T_c$ cuprates as long as the induced gap is dominantly $s$-wave. 
In fact Wang {\it et al.} \cite{WDF+13} observed an isotropic  gap opening on the Dirac branch on a thin film of \bise{} on a $\mathrm{Bi_2 Sr_2 Ca Cu_2 O_{8+\delta}}$ substrate below the superconducting transition temperature. 
While the mechanism for the larger value of the inferred surface-state gap compared to the bulk gap in Ref.~\onlinecite{WDF+13} remains unknown~\cite{LCY14}  and the results of Ref.~\onlinecite{WDF+13} have not been reproduced to date~\cite{YPW+14}, our results should apply as long as the induced isotropic gap is dominantly $s$-wave.

The setup of \bise{} proximity coupled to superconducting NbSe$_{2}$ recently studied using ARPES and point-contact transport in Ref.~\onlinecite{XAB+14} actually satisfy the condition of TM-SC structure as defined in this paper, according to their spin-momentum locking observations.
Our results imply that the same system can support Majorana bound states at vortex cores with spatial separation between the top (naked) surface Majorana and the bottom (buried) surface Majorana.
So far little attention has been given to experimentally distinguishing the two surfaces of TI in such a heterostructure, although Ref.~\onlinecite{XAB+14} showed how the spectral gap at the Dirac point depends on the film thickness presumably due to varying degrees of coupling between the two surfaces.    
One way to experimentally identify the surface would be to use ARPES and look for the normal-state Fermi surface of the substrate.
The Dirac state signal probed simultaneously with the substrate will be coming from both the top surface and the interface.
When the film is thick enough to not show the substrate Fermi surface, the Dirac state signal will be coming from the naked top surface.
In order to test our predictions we propose in-field STM measurements looking for Majorana bound states in a TM-SC setup like that of Ref.~\onlinecite{XAB+14} in which spin-momentum locking is confirmed, with further attention given to distinguishing signals from each surface.


\noindent
{\bf Acknowledgements}. We thank Z. Hasan for discussions that motivated the work and H. Yao for useful discussions. K. Lee, A. Vaezi, and E.-A.K. were supported in part by NSF CAREER Grant No. DMR-0955822. M.H.F. and E.-A.K. were supported in part by NSF Grant No. DMR-1120296 to the Cornell Center for Materials Research.

%

\end{document}